\documentclass[conference,10pt]{IEEEtran}
\usepackage{graphicx,cite,epsfig,amssymb,amsmath,subfigure,url,stfloats,latexsym}
\usepackage{bm}
\usepackage{algorithm}
\usepackage{algorithmic}
\usepackage{cases}
\usepackage{citesort}
\usepackage{color}
\usepackage{epstopdf}
\usepackage{amssymb}
\usepackage{cite}
\usepackage[caption=false,font=normalsize,labelfont=sf,textfont=sf]{subfig}
\usepackage{amsmath,lipsum}
\usepackage{cuted}

\newtheorem{theorem}{Theorem}

\newtheorem{lemma}{Lemma}

\newtheorem{corollary}{Corollary}

\begin{document}
	
	\title{Parameter Design for Secure Affine Frequency Division Multiplexing Waveform}
	
	 \author{\IEEEauthorblockN{Di Zhang\IEEEauthorrefmark{1}, Zeyin Wang\IEEEauthorrefmark{1}, Yanqun Tang\IEEEauthorrefmark{2}, Dongdong Wu\IEEEauthorrefmark{1} and Muzi Yuan\IEEEauthorrefmark{3}}
	 	\IEEEauthorblockA{\IEEEauthorrefmark{1}School of Electrical and Information Engineering, Zhengzhou University, Zhengzhou 450001, China}
	 	\IEEEauthorblockA{\IEEEauthorrefmark{2}School of Electronics and Communication Engineering, Sun Yat-sen University, Shenzhen 518107, China}
	 		 \IEEEauthorblockA{\IEEEauthorrefmark{3}College of Electronic Science and Technology, National University of Defense Technology, Changsha 410073, China\\
	 		 	 Email: dr.di.zhang@ieee.org, \{eiezeyinwang, eieddwu\}@gs.zzu.edu.cn, tangyq8@mail.sysu.edu.cn, ymz@nudt.edu.cn}}

	\maketitle 
	\pagestyle{empty}
	\thispagestyle{empty}
	\begin{abstract}
		The secure affine frequency division multiplexing (AFDM) waveform design is a main concern in high-mobility networks. In this article, we employ the four key parameters in AFDM to design secure waveforms, and afterward we analyze the role of the four parameters to reveal the design guideline. We find that $c_1$ is bounded by the Doppler shifts and preset guard, i.e., $c_1$ needs to be the discrete value given by ${\frac{2\alpha^C_\text{max}+1}{2N}{\leq}{c_1=\frac{2k+1}{2N}}{\leq}\frac{2\alpha_\text{max}+1}{2N}, k\in{\mathbb{N}^+}}$. The parameter $c_2$ exhibits a minimum periodicity of $1$, with the effective range $[0, 1]$ rather than any real number. The adjustable parameters $c_1$ and $c_2$ introduce additional degrees of freedom to the AFDM waveform, thereby enhancing the anti-eavesdropping performance. In addition, excessive $L_\text{max}$ that determines the preset guard interval leads to a security-risk interval and poses eavesdropping risks. Therefore, the optimal $L_\text{max}$ equals maximum delay. Numerical simulations verify the accuracy of our analysis.
	\end{abstract}
		

	\IEEEpeerreviewmaketitle

	\section{Introduction}	
	The rapid evolution of wireless networks toward the sixth generation is anticipated to support extensive high-mobility and fast time-varying scenarios, such as high-speed railways, unmanned aerial vehicle (UAV) networks, vehicular networks, and low Earth orbit satellites networks \cite{Xuewan_Chinacomm,Xuewan_IoTJ}. These high-mobility and fast time-varying characteristics endow the channel with time-selective and frequency-selective fading properties, which severely degrades the performance of conventional orthogonal frequency division multiplexing (OFDM) due to inter-carrier interference \cite{double_selective_channel,degrading_OFDM}. To address this issue, a new waveform called affine frequency division multiplexing (AFDM), based on the discrete affine Fourier transform (DAFT), has been proposed, showing superior delay-Doppler-separating performance \cite{AFDM1,AFDM2,Haoran2}. Besides, AFDM can achieve low pilot overhead, low complexity and backward compatibility, positioning it as a promising candidate for high-mobility communications \cite{AFDM1,AFDM2,Haoran2}. 
    The AFDM can also be applied in various scenarios, and there are some prior studies, e.g., integrated sensing and communications (ISAC) \cite{ISAC1,JiajunISAC2,ISAC3}, pilot design \cite{HaoranPilot}, multiple-input multiple-output (MIMO) \cite{HaoranMIMO}, and sparse code multiple access (SCMA) \cite{Luoqu}. 
	
	On the other hand, the broadcast nature of the wireless medium makes intrinsic anti-eavesdropping performance the crucial issue in AFDM waveform design \cite{Haibo,wanghuiming1}. 
	Besides, due to the superior delay-Doppler-separating performance in combating complex channel conditions, AFDM waveform contains rich private information such as channel state, velocity, or direction, which is more likely to pose severe eavesdropping risks \cite{JiajunISAC2,wanghuiming2}.
	However, to the best of our knowledge, the related intrinsic anti-eavesdropping analyses remain scarce and admissible ranges of parameters remain unclear, which may lead to insecure parameter configurations in practical deployments, thereby exposing the AFDM waveform to potential privacy leakage or eavesdropping risks.
	
	Motivated by the above discussions, in this article, we analyze the impact of the key parameters $c_1,c_2,\alpha_{\text{max}}$ and $L_\text{max}$ on AFDM waveform. Furthermore, we provide design guidelines for each parameter as follows. There is an admissible range for the discrete value parameter $c_1=\frac{2k+1}{2N}\in[\frac{2\alpha^C_\text{max}+1}{2N},\frac{2\alpha_\text{max}+1}{2N}], k\in{\mathbb{N}^+}$, while the periodicity of $c_2$ is $1$ and the efficient range of $c_2$ is $[0, 1]$. The adjustable parameters $c_1$ and $c_2$ introduce two additional degrees of freedom to the AFDM waveform, enhancing the anti-eavesdropping performance. The optimal value for $L_\text{max}$ equals $L^{\mathrm{C}}_{\text{max}}$. 
	
	The rest of the article is organized as follows. The considered system is elaborated in Section II. Section III presents the analysis and design guidelines for the parameter. The simulation results are presented in Section IV, and the article is finally concluded in Section V.
	\section{AFDM-based System model}
	
	In this section, we review the basic principles of the AFDM-based system. As shown in Fig.~1, let $\mathbf{x}$ denote a vector of $N$ quadrature amplitude modulation (QAM) symbols on the DAFT domain.\vspace{-5pt}
	\begin{figure}[h]
		\centering
		\includegraphics[width=2.4in]{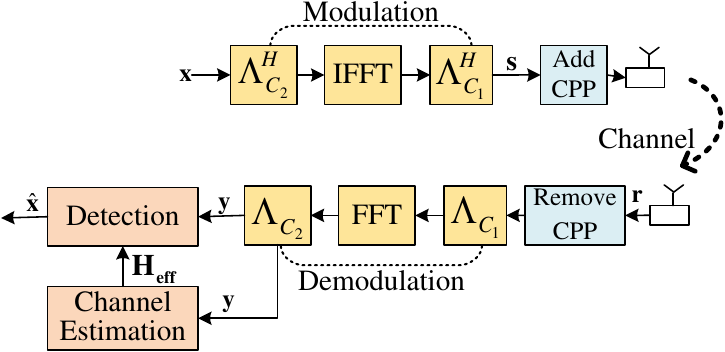}
		\caption{AFDM-based system block diagram.}
		\label{fig:secure_communication_system_model}
	\end{figure}\vspace{-3pt}
	The $N$ points inverse DAFT (IDAFT) modulation is performed to map $\mathbf{x}$ into the time domain as
	\begin{equation}
	s_n=\frac{1}{\sqrt{N}}\sum_{m=0}^{N-1}x_me^{j2\pi(c_1n^2+\frac{mn}{N}+c_2m^2)},
	\end{equation}
	where $c_1$ and $c_2$ are key parameters in AFDM waveform, $N$ equals the number of the symbols, ${n=0,1,\cdots, N-1}$ and ${m=0,1,\cdots, N-1}$ stand for indexes of signals in the time domain and DAFT domain, respectively. In matrix form, (1) will be $\mathbf{s}=\mathbf{A^\mathrm{H}x}=\mathbf{\Lambda}_{c_1}^{\mathrm{H}}\mathbf{F^\mathrm{H}}\mathbf{\Lambda}_{c_2}^{\mathrm{H}}\mathbf{x}$, where $\mathbf{\Lambda}_{c_1}=\text{diag}(e^{-j2\pi{c_1}n^2})$, $\mathbf{\Lambda}_{c_2}=\text{diag}(e^{-j2\pi{c_2}m^2})$, and $\mathbf{F}$ stands for the discrete Fourier transform (DFT) matrix. To combat the multipath propagation and enable the periodicity of symbols, an $L_{cp}$-long chirp-periodic prefix (CPP) is used, where $L_{cp}$ is any integer greater than or equal to the maximum delay in multipath channels. The prefix can be expressed as
		\begin{equation}
		s_n=S_{N+n}e^{-j2\pi{c_1}(N^2+2Nn)},\quad n=-L_{cp},\cdots,-1.
		\end{equation}
After transmission over the channel, the received symbol is
\begin{equation}
	r_n=\sum_{L^C=0}^{\infty}s({n-L^C})g_n(L^C)+w_n,
\end{equation}
where $w_n\sim\mathcal{CN}(0,N_0)$ is an additive Gaussian noise, and
\begin{equation}
	g_n(l)=\sum_{i=1}^{P}h_ie^{-j2{\pi}f_in}\delta(l-L^C_i)
\end{equation}
is the channel impulse response at time $n$ and delay $l$, where $P$ represents the number of paths, and $h_i$, $L^C_i$ and $f_i$ stand for the complex gain in channel, delay, and the Doppler shift in the $i$-th path, respectively. 
we can write (3) in the matrix form as
\begin{equation}
	\mathbf{r}=\mathbf{Hr}+\mathbf{w},
\end{equation}
where $\mathbf{w}\sim\mathcal{CN}(\mathbf{0},N_0\mathbf{I})$, $\mathbf{H}=\sum_{i=1}^{P}h_i\mathbf{\Gamma}_{\textnormal{CPP}_i}\mathbf{\Delta}_{f_i}\mathbf{\Pi}^{L^C_i}$, $\mathbf{\Delta}_{f_i}=\textnormal{diag}(e^{-j2{\pi}f_in},n=0,1,\cdots,N-1)$, $\mathbf{\Pi}$ is the forward cyclic-shift matrix, and $\mathbf{\Gamma}_{\textnormal{CPP}_i}$ is a $N{\times}N$ diagonal matrix, i.e., 
\begin{equation}
	\mathbf{\Gamma}_{\textnormal{CPP}_i}=\textnormal{diag}(	
	\begin{cases}
		e^{-j2{\pi}c_1(N^2-2N(L^C_i-n))}        \   n < {L^C_i}  \\
		1  \quad          \qquad  \qquad\	\qquad\qquad	n\geq L^C_i 
	\end{cases}	).
\end{equation}
After receiving the symbols, discarding the CPP and $N$ points DAFT are performed to map the time domain symbols $\mathbf{r}$ into the DAFT domain symbols $\mathbf{y}$ as
\begin{equation}
	y_m=\frac{1}{\sqrt{N}}\sum_{n=0}^{N-1}r_ne^{-j2\pi(c_1n^2+\frac{mn}{N}+c_2m^2)}.
\end{equation}
Equation (7) also can be expressed by the matrix as
\begin{equation}
	\mathbf{y}=\sum_{i=1}^{P}h_i\mathbf{A}\mathbf{\Gamma}_{\textnormal{CPP}_i}\mathbf{\Delta}_{f_i}\mathbf{\Pi}^{l_i}\mathbf{A^\mathrm{H}x}+\mathbf{Aw}=\mathbf{H}_{\textnormal{eff}}\mathbf{x}+\tilde{\mathbf{w}},
\end{equation}
where $\mathbf{H}_{\textnormal{eff}}=\mathbf{A}\mathbf{H}\mathbf{A^\mathrm{H}}$ and $\tilde{\mathbf{w}}=\mathbf{\Lambda}_{c_2}\mathbf{F}\mathbf{\Lambda}_{c_1}\mathbf{w}$.
The input-output relation of AFDM can be written as
\begin{equation}
	y_m=\sum_{i=1}^{P}h_ie^{j\frac{2\pi}{N}(Nc_1{L^C_i}^2-qL^C_i+Nc_2(q^2-m^2))}x_q+\tilde{w}_m,
\end{equation}
where $q=(m+\text{loc}_i)_N$ and $\text{loc}_i=(\alpha_i+2Nc_1L^C_i)_N$.
	\section{Parameter Analysis And Design}
	In the AFDM-based system, accurate demodulation and channel estimation are critical to detect the information. Besides, there are 4 key parameters (i.e., $c_1, c_2, \alpha_\text{max}, L_\text{max}$) playing great roles in the demodulation and channel estimation, where $\alpha_\text{max}$ and $L_\text{max}$ represent the preset Doppler shift and delay guard interval between transceivers to resist the maximum Doppler shift of $\alpha_\text{max}$ and the maximum delay of $L_\text{max}$ in actual channels, respectively. Therefore, in this section, we will analyze the impact of various parameters on the AFDM waveform and ultimately reveal guidelines for parameter designs for secure AFDM. To isolate the impact of the target parameter, all non-target parameters are ideal during analysis and simulation. It is evident that the accuracy of channel estimation depends on the input-output relation of the system, which will be discussed in detail next.
	We can rewrite (8) as
	\begin{equation}
		\mathbf{y}=\sum_{i=1}^{P}h_i\mathbf{H}_{i}\mathbf{x}+\tilde{\mathbf{w}},
	\end{equation}
	where $\mathbf{H}_{i}=\mathbf{A}\mathbf{\Gamma}_{\textnormal{CPP}_i}\mathbf{\Delta}_{f_i}\mathbf{\Pi}^{l_i}\mathbf{A^\mathrm{H}}$. Because the input and output signals are all in the DAFT domain, $p$ and $q$ denote the index of the input signal and the output signal, respectively. It can be proven that
	$H_i[p,q]$ can be given as
	\begin{equation}
		H_i(p,q)=\frac{1}{N}e^{j\frac{2\pi}{N}(Nc_1{L^C_i}^2-qL^C_i+Nc_2(q^2-p^2))}\mathcal{F}_i(p,q),
	\end{equation}
	where $\mathcal{F}_i(p,q)$ is
	\begin{equation}
		\begin{aligned}
			\mathcal{F}_i(p,q) &=\sum_{n=0}^{N-1}e^{-j\frac{2\pi}{N}(p-q+\alpha_i+2Nc_1L^C_i)}  \\
			&=\frac{e^{-j2\pi(p-q+\alpha_i+2Nc_1L^C_i)}-1}{e^{-j\frac{2\pi}{N}(p-q+\alpha_i+2Nc_1L^C_i)}-1}.
		\end{aligned}
	\end{equation}
	Besides, (12) can be further represented as
	\begin{equation}
		\mathcal{F}_i(p,q)=	
		\begin{cases}
			N      \qquad   q=p+\alpha_i+(2\alpha_{c_1}+1){L^C_i},  \\
			0  \quad          \qquad  	\text{otherwise},
		\end{cases}
	\end{equation}
	where $\frac{2Nc_1-1}{2}\triangleq\alpha_{c_1}, \alpha_{c_1}\in \mathbb{N}^+$ represented the maximum ability to separate Doppler shifts, which is determined by $c_1$. As we can see, the value of $\mathcal{F}_i(p,q)$ is determined by the delay $L^C_i$, Doppler shift $\alpha_i$, and $\alpha_{c_1}$. By substituting (13) into (11), $H_i[p,q]$ can be expressed as
	
	\begin{equation}
		H_i(p,q)=	
		\begin{cases}
			e^{j\frac{2\pi}{N}t}\quad \ q=p+\alpha_i+(2\alpha_{c_1}+1){L^C_i},  \\
			0   \qquad\qquad\qquad	\text{otherwise},
		\end{cases}		
	\end{equation}
	where $t=Nc_1{L^C_i}^2-qL^C_i+Nc_2(q^2-p^2)$. Therefore, the input-output relation can be rewritten as
	\begin{equation}
		y_p=\sum_{i=1}^{P}h_ie^{j\frac{2\pi}{N}(Nc_1{L^C_i}^2-qL^C_i+Nc_2(q^2-p^2))}x_q+\tilde{w}_p,
	\end{equation}
	where $q=p+\alpha_i+(2\alpha_{c_1}+1){L^C_i}$. Therefore, AFDM is capable of separating the propagation paths with distinct delays or Doppler shifts in the one-dimensional DAFT domain because of the input-output relation. In this case, the value of $c_1$ needs to satisfy $\alpha_{c_1}=\frac{2Nc_1-1}{2}\in \mathbb{N}^+$.The demodulated frame structure is depicted as Fig.~2, in which the separated pilots are located in an interval $[-\alpha_{c_1},+\alpha_{c_1}]$ with a length of $(2\alpha_{c_1}+1)$, and arranged in the order of $[L^C_\text{max},L^C_\text{max-1},\cdots,L^C_0]$.\vspace{-5pt}
	\begin{figure}[h]
		\centering
		\includegraphics[height=1.4cm,width=8.34cm]{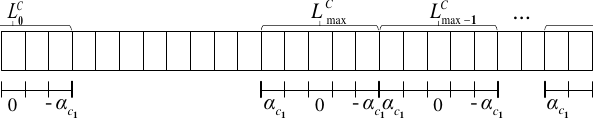}
		\caption{The demodulated AFDM frame structure.}
		\label{fig The demodulated AFDM frame structure}
	\end{figure}\vspace{-10pt}
	\begin{figure}[h]
	\centering
	\includegraphics[width=3.4in]{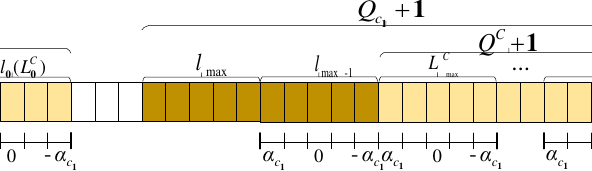}
	\caption{The estimation structure for demodulated AFDM signals.}
	\label{fig:The estimation structure of signals}
	\end{figure}	\vspace{-5pt}
\subsection{$L_\text{max}$ Design}\label{CA}
	To ensure progressive analysis and readability, we begin with the analysis of $L_\text{max}$. Through demodulation, the receiver separates the Doppler and delay components, partitioning them into $L^C_\text{max}$ ordered blocks. In the subsequent channel estimation, the parameter $l_\text{max}$ determines the ability to detect the maximum delay, and the delay blocks to be estimated are ordered in descending sequence $[l_\text{max},l_\text{max-1},\cdots,l_0]$, which is shown as Fig.~3.
The actual shifted pilots through multipath are confined to the interval
	\begin{equation}
	[-\alpha^C_\text{max}+(2\alpha_{c_1}+1)L^C_0,\alpha^C_\text{max}+(2\alpha_{c_1}+1)L^C_\text{max}],
\end{equation}
 where $L^C_\text{max}$ and $\alpha^C_\text{max}$ stand for the actual maximum delay and Doppler shift in multipath, respectively. However, the interval of symbols to be estimated is
	\begin{equation}
	[-\alpha_{c_1}+(2\alpha_{c_1}+1)l_0,\alpha_{c_1}+(2\alpha_{c_1}+1)l_\text{max}].
\end{equation}
When $l_\text{max}\geq{L}^C_\text{max}$, the detected range fully covers the practical pilot-shifted ranges, which enables the complete channel estimation. Conversely, if $l_\text{max}<L^C_\text{max}$, the actual multipath channel is separated into $L^C_\text{max}$ blocks, exceeding the receiver’s $l_\text{max}$ blocks in channel estimation, which leads to the missed detection of delays within the interval $(l_\text{max}, L^C_\text{max}]$.

On the other hand, when $l_\text{max}>L_\text{max}$, the interval length of symbols to be estimated is $(2\alpha_{c_1}+1)(l_\text{max}+1)$. However, the preset guard interval in the frame is
 	\begin{equation}
	[-\alpha_\text{max}+(2\alpha_\text{max}+1)L_0,\alpha_\text{max}+(2\alpha_\text{max}+1)L_\text{max}]
\end{equation}
with the length of $(2\alpha_\text{max}+1)(L_\text{max}+1)$, which is shorter than the detected interval length $(2\alpha_{c_1}+1)(l_\text{max}+1)$ in channel estimation. Therefore, when $l_\text{max}>L_\text{max}$, the data beyond the guard interval will be falsely detected as spurious paths.

Therefore, when $l_\text{max}<L^C_\text{max}$ or $l_\text{max}>L_\text{max}$, the channel estimation becomes inaccurate, degrading communication performance. When $L^C_\text{max}\leq{l_\text{max}}\leq{L}_\text{max}$, all delays in multipath are detected and complete delay estimation is achieved. On the other hand, since $l_\text{max}$ is the parameter for the receiver’s channel estimations and does not affect the AFDM waveform, the above analysis equally applies to eavesdroppers. Therefore, the preset $L_\text{max}$ in transceivers introduces a security-risk range $[L^C_\text{max},L_\text{max}]$. If the eavesdropper’s parameter configuration resides within the interval, it can fully estimate all delays in multipath, posing eavesdropping risks. Consequently, the design of $L_\text{max}$ for legitimate parties needs to minimize this security-risk range by ensuring $L_\text{max}$ aligns with the actual maximum channel delay $L^C_\text{max}$, i.e., $L^C_\text{max}=l_\text{max}=L_\text{max}$.
\vspace{-5pt}
\subsection{$c_1$ and $c_2$ Designs}\label{CB}\vspace{-10pt}
		\begin{figure}[h]
		\centering
		\includegraphics[width=3.4in]{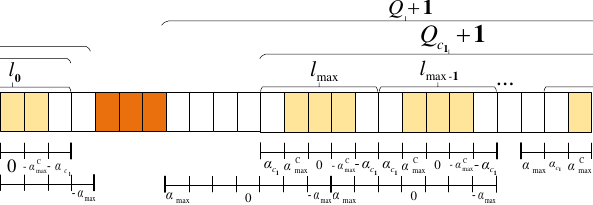}
		\caption{The structure of demodulated AFDM signals.}
		\label{The structure of demodulated AFDM signals}
	\end{figure}
	Due to $c_1=\frac{2\alpha_{c_1}+1}{2N}$, we will first provide the range for $c_1$ from the perspective of input-output relation, which separates the Doppler and delay components. Through demodulation, the shifted pilot is separated into $L^C_\text{max}$ sequentially ordered blocks, each further sorted by their Doppler shifts. The Fig.~4 shows the case for $\alpha^C_\text{max}\leq\alpha_{c_1}\leq\alpha_\text{max}$. The actual shifted pilot index $p$ of the $i$-th path is confined to
	\begin{equation}
		-\alpha^C_\text{max}+(2\alpha^C_\text{max}+1)L^C_i\leq{p}\leq\alpha^C_\text{max}+(2\alpha^C_\text{max}+1)L^C_i.
	\end{equation}
However, the range of separated pilot by demodulation in the $i$-th path is
		\begin{equation}
		-\alpha_\text{max}+(2\alpha_{c_1}+1)L^C_i\leq{p}\leq\alpha^C_\text{max}+(2\alpha_{c_1}+1)L^C_i.
	\end{equation}
	Because of $\alpha_{c_1}\geq\alpha^C_\text{max}$, the pilot-separated range by demodulation fully covers the pilot-shifted range and the Doppler shift of each path can be estimated.
	Conversely, if $\alpha_{c_1}<\alpha^C_\text{max}$, the length of Doppler-shifted sub-block for each path is $2\alpha^C_\text{max}+1$, exceeding the receiver’s $2\alpha_{c_1}+1$ block length separated by demodulation and detected in channel estimation. This results in missed detection of Doppler shifts within interval $(\alpha_{c_1},\alpha^C_\text{max}]$. Furthermore, owing to the linear mapping between separable Doppler shifts and delays, the undetected Doppler components accumulate in the adjacent (i.e., $l_{i-1}$-th or $l_{i+1}$-th) delay block detection, causing simultaneous errors in Doppler shift and delay estimations.
	
	On the other hand, when $\alpha_{c_1}\leq\alpha_\text{max}$, the interval of the detected symbols in channel estimation is
	\begin{equation}
		[-\alpha_{c_1}+(2\alpha_{c_1}+1)l_0,\alpha_{c_1}+(2\alpha_{c_1}+1)l_\text{max}]
	\end{equation}
	with the length of $(2\alpha_{c_1}+1)(l_\text{max}+1)$. Besides, the preset guard interval between the transmitter and receiver is
	\begin{equation}
		[-\alpha_\text{max}+(2\alpha_\text{max}+1)L_0,\alpha_\text{max}+(2\alpha_\text{max}+1)L_\text{max}]
	\end{equation}
	with the length of $(2\alpha_\text{max}+1)(L_\text{max}+1)$, which covers the estimation interval and enables a complete Doppler shift estimation. Otherwise, if $\alpha_{c_1}>\alpha_\text{max}$, the estimation for separated shifted pilots will be interfered with by the data beyond the guard interval, i.e., in channel estimation, the range to be estimated may exceed the preset guard interval, leading to misjudgment of the data portion as spurious paths.
	Therefore, for the AFDM transceivers, the admissible range of parameter $\alpha_{c_1}$ without degrading the bit error rate (BER) performance is 
			\begin{equation}
	\alpha^C_\text{max}\leq\alpha_{c_1}\leq\alpha_\text{max},\alpha_{c_1}=\frac{2Nc_1-1}{2}\in{\mathbb{N}^+}.
			\end{equation}
			 Therefore, the admissible discrete values for $c_1$ is
		\begin{equation}
		\frac{2\alpha^C_\text{max}+1}{2N}\leq{c_1}=\frac{2\alpha_{c_1}+1}{2N}\leq\frac{2\alpha_\text{max}+1}{2N},\alpha_{c_1}\in{\mathbb{N}^+}.
	\end{equation}
	 
	 On the other hand, from the perspective of modulation and demodulation, reviewing equations (1) and (7), key parameters $c_1$ and $c_2$ reside in complex exponential terms and exhibit inherent periodicity. Equation (1) can be rewritten as
			\begin{equation}
		\begin{aligned}
		s_n&=\frac{1}{\sqrt{N}}\sum_{m=0}^{N-1}x_me^{j2\pi(c_1n^2+\frac{mn}{N}+c_2m^2)}\\
		&=\frac{1}{\sqrt{N}}e^{j2\pi{c_1}n^2}\sum_{m=0}^{N-1}x_me^{j2\pi(\frac{mn}{N}+c_2m^2)}.
		\end{aligned}
	\end{equation}
	Given the periodicity of $e^{jx}$, parameter $c_1$ inherits a periodicity satisfying $e^{2\pi{T_{c_1}}n^2}=e^{2k\pi},k\in{\mathbb{Z}},n=1,2,\cdots,N$, i.e., $T_{c_1}=\frac{k}{n^2}$. Therefore, the minimal positive period of $c_1$ is the least common multiple $1$. Similarly, parameter $c_2$ also inherits a periodicity satisfying $e^{2\pi{T_{c_2}}m^2}=e^{2k\pi},m=1,2,\cdots,N$, i.e., $T_{c_2}=\frac{k}{m^2}$. The minimal positive period of $c_2$ also is the least common multiple $1$. Similarly, we can also obtain the minimum positive period $1$ of $c_1$ and $c_2$ in demodulation.
	
	Summing up the above analysis, parameter $c_2$ exhibits a minimum periodicity of $1$, which indicates that the effective range for $c_2$ is $[0, 1]$, rather than any real number in $R$ in \cite{AFDM1,HaoranPilot,JiajunISAC2}. Although $c_1$ is periodic in modulation and demodulation, as mentioned previously, $c_1$ also has an important impact on estimating Doppler frequency shift and needs to be the discrete value. Therefore, the admissible value for $c_1$ is
	\begin{equation} 
		\frac{2\alpha^C_\text{max}+1}{2N}\leq{c_1}=\frac{2\alpha_{c_1}+1}{2N}\leq\frac{2\alpha_\text{max}+1}{2N},\alpha_{c_1}\in{\mathbb{N}^+}, 
	\end{equation}
	rather than the fixed default value $c_1=\frac{2\alpha_\text{max}+1}{2N}$ as in many works \cite{HaoranPilot,JiajunISAC2,AFDM1}. The adjustable parameters $c_1$ and $c_2$ introduce additional degrees of freedom to the AFDM waveform, enhancing the anti-eavesdropping performance.
	
	\section{Simulation Results}
	In this section, the validity of the analysis is verified through numerical results. The impacts of the parameters on secure AFDM designs are also demonstrated. The Doppler shift and delay in each path are random, of which the maximum delay $L^C_\text{max}$ and the maximum Doppler shift $\alpha^C_\text{max}$ are $5$ and $3$, respectively. The energy ratio of the pilot to noise is $40$ dB and the number of symbols $N$ equals $512$.

	\begin{figure}[h]
	\centering
	\includegraphics[height=4.4cm,width=4.34cm]{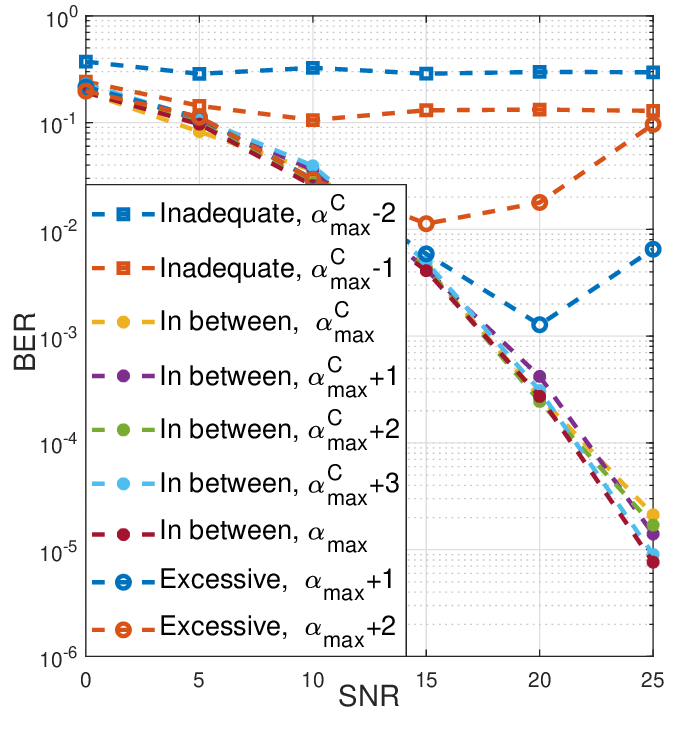}
	\includegraphics[height=4.4cm,width=4.34cm]{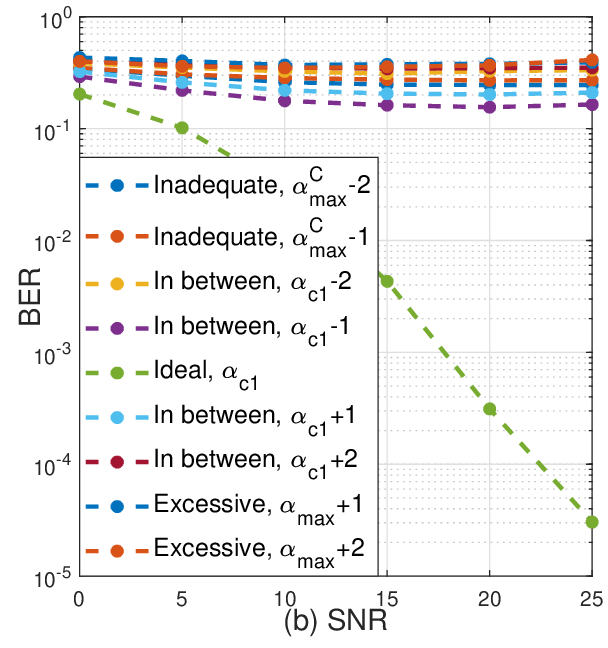}\vspace{-5pt}
	\caption{(a) BER versus SNR when different $\alpha_{c_1}$ are applied by transmitter and receiver. (b) Eavesdropper's BER versus SNR with different $\alpha_{c_1}$.} \vspace{-15pt}
	\label{fig:2}
	\end{figure}
	
Fig.~5(a) shows the BER performance versus the signal-to-noise ratio (SNR) when different $\alpha_{c_1}$ are applied by transmitter and receiver, while Fig.~5(b) shows the eavesdropper's BER performance versus SNR with different $\alpha_{c_1}$ at eavesdropper. The preset parameters $\alpha_\text{max}$ equals 7, representing the guard interval to resist the maximum Doppler shift. To isolate the impact of parameters $\alpha_{c_1}$, we set other parameters are ideal, i.e., $L^C_\text{max}=l_\text{max}=L_\text{max}$ and receiver's $c_2$ is consistent with the transmitter's. In Fig.~5(a), for the legitimate receiver, the BER performance remains consistent and superior when $\alpha_{c_1}\in[\alpha^C_\text{max},\alpha_\text{max}]$, significantly outperforming cases where $\alpha_{c_1}>\alpha_\text{max}$ and $\alpha_{c_1}<\alpha^C_\text{max}$, which demonstrates our analysis that $\alpha_\text{max}$ provides an admissible range for parameter $\alpha_{c_1}$ and $c_1$ without degrading the BER performance. 
When $\alpha_{c_1}<\alpha^C_\text{max}$, the performance severely deteriorates. This is because the system with small $\alpha_{c_1}$ fails to resolve all Doppler shifts, causing beyond $\alpha_{c_1}$ Doppler shifts to accumulate into adjacent delay blocks. Due to the linear arrangement of Doppler-delay symbols, this accumulating introduces simultaneous errors in delay and Doppler estimations, ultimately degrading the BER. 
On the other hand, when $\alpha_{c_1}>\alpha_\text{max}$, the performance also severely deteriorates. This is because $\alpha_{c_1}$ exceeding $\alpha_\text{max}$ extends the channel estimation region beyond the preset guard interval, falsely detecting data segments as spurious paths. In addition, we notice higher SNR also exacerbates BER performance. This is because given the fixed 40 dB pilot, increased data SNR amplifies interference from data-induced false paths during channel estimation, thereby raising the BER.
As shown in Fig.~5(b), eavesdropping risks arise only when the eavesdropper’s $\alpha_{c_1}$ matches the transmitter’s, which validates the anti-eavesdropping performance for $c_1$.

\begin{figure}[h]
	\centering
	\includegraphics[width=2.5in]{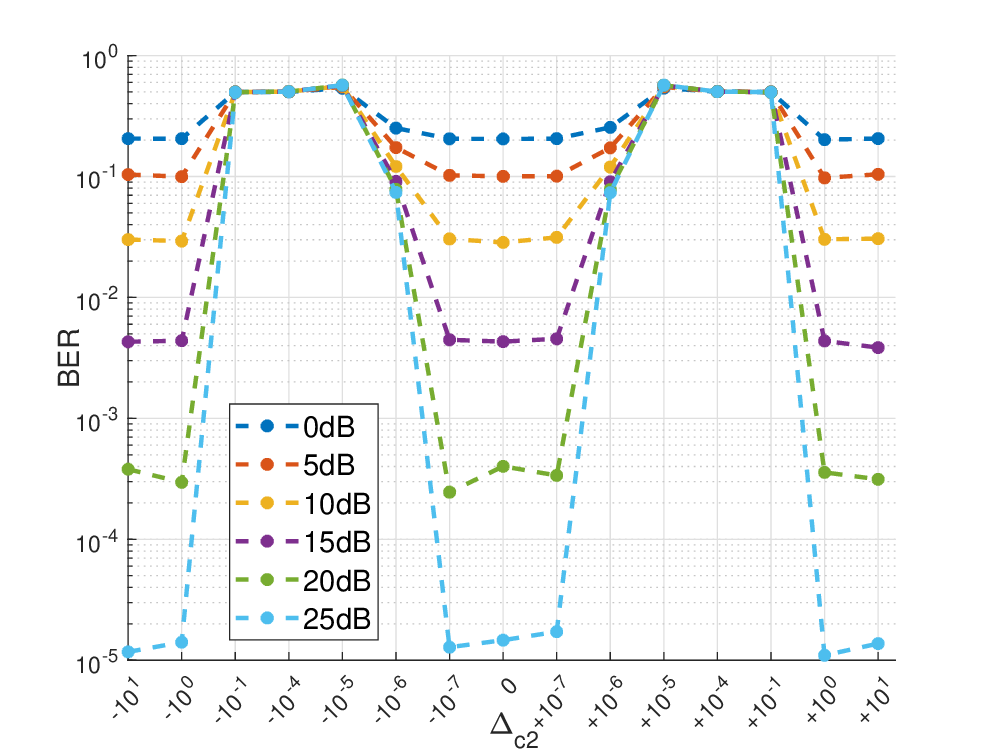}\vspace{-5pt}
	\caption{Eavesdropper's BER versus the value of $c_2$ with different SNR.}\vspace{-10pt}
	\label{fig:3}
\end{figure}

Fig.~6 illustrates the eavesdropper's BER performance versus the deviation of the correct $c_2$. To isolate the impact of $c_2$, we set $\alpha^C_\text{max}=\alpha_{c_1}=\alpha_\text{max}$, $L^C_\text{max}=l_\text{max}=L_\text{max}$, and $c_1$ identical to the transmitter's. In Fig.~6, the impact of deviation of $c_2$ on BER performance is negligible when the deviation is below $10^{-7}$. However, BER performance degrades rapidly to $10\%$ at deviation of $10^{-6}$ and becomes $50\%$ at deviation of $10^{-5}$, revealing the superior anti-eavesdropping performance for $c_2$. Notably, for deviations of $\pm1$ or $\pm10$, the BER performance is consistent with the deviation-free case, which demonstrates our analysis that $c_2$ exhibits periodicity of $1$, i.e., the effective range of $c_2$ is confined to $[0, 1]$.  
\vspace{-5pt}
	\begin{figure}[h]
	\centering
	\includegraphics[width=2.5in]{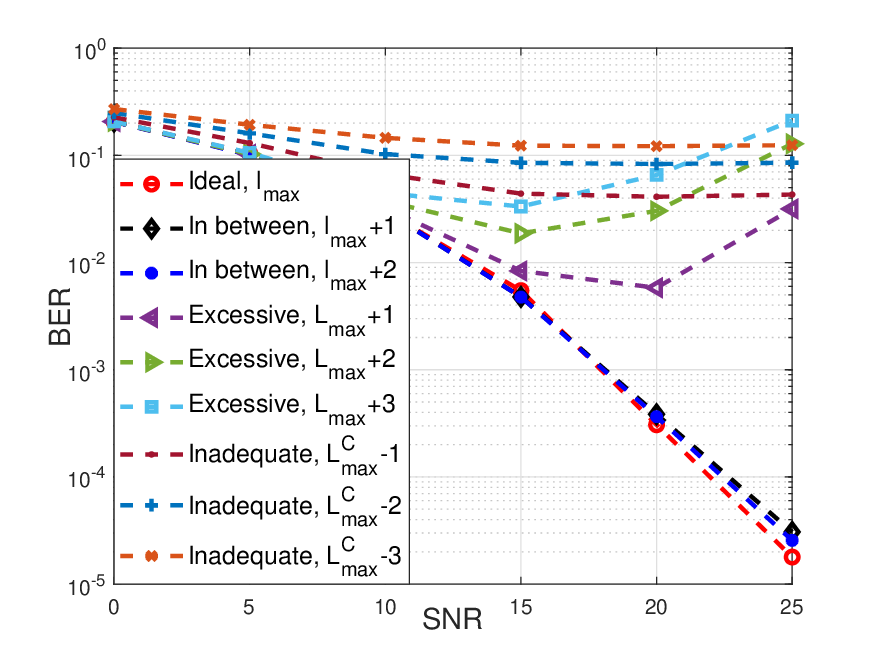}\vspace{-5pt}
	\caption{Eavesdropper's BER versus SNR with different $l_\text{max}$.} \vspace{-5pt}
	\label{fig:5}
\end{figure}
Fig.~7 shows the eavesdropper's BER performances versus SNR given different $l_\text{max}$. The preset parameters $L_\text{max}$ equals 7. To isolate the impact of parameters $l_\text{max}$, non-target parameters are as follows, i.e., $\alpha^C_\text{max}=\alpha_{c_1}=\alpha_\text{max}$, and eavesdropper's $c_1$ and $c_2$ matching the transmitter’s.
The eavesdropper's performance remains consistent and superior when $l_\text{max}\in[L^C_\text{max},L_\text{max}]$, significantly outperforming cases where $l_\text{max}<L^C_\text{max}$ and $l_\text{max}>L_\text{max}$, which demonstrates our analysis that a large $L_\text{max}$ in transceivers introduces a security-risk range $[L^C_\text{max},L_\text{max}]$.
When $l_\text{max}<L^C_\text{max}$, the performance severely deteriorates. This is because the eavesdropper with small $l_\text{max}$ fails to estimate all delays, ultimately degrading the BER. 
On the other hand, when $l_\text{max}>L_\text{max}$, the performance also severely deteriorates. This is because $l_\text{max}$ exceeding $L_\text{max}$ also extends the channel estimation region beyond the preset guard interval, falsely detecting data segments as spurious paths. In addition, higher SNR also exacerbates BER performance. This is because given the fixed 40 dB pilot, increased data SNR amplifies interference from data-induced spurious paths during channel estimation, thereby raising the BER.

	\section{Conclusion} 
	In this work, we addressed the parameter design for secure AFDM waveform by analyzing four key parameters: $c_1, c_2, \alpha_\text{max}$, and $L_\text{max}$. We demonstrated that $c_1=\frac{2k+1}{2N}, k\in{\mathbb{N}^+}$ needs to be adopted as the discrete value from the interval $[\frac{2\alpha^C_\text{max}+1}{2N},\frac{2\alpha_\text{max}+1}{2N}]$, while the periodicity of $c_2$ is $1$ and the efficient range of $c_2$ is $[0, 1]$. The adjustable parameters $c_1$ and $c_2$ introduce additional degrees of freedom to the AFDM waveform, enhancing the anti-eavesdropping performance. In addition, the anti-eavesdropping performance will reduce if $L_{\text{max}}$ exceeds $L^{\mathrm{C}}_{\text{max}}$. Therefore, the optimal value for $L_\text{max}$ equals $L^{\mathrm{C}}_{\text{max}}$. Numerical results validated these insights. Our work provides foundational design guidelines for secure AFDM waveform.
 
	\bibliographystyle{IEEEtran}
	\bibliography{ices}
	
\end{document}